\documentclass[structabstract]{aa}
\usepackage{graphicx,amsfonts,amssymb,txfonts,natbib,latexsym}
\newcommand{\plotwd}{8.2cm}
\newcommand{\plotwdtwo}{16.4cm}
\newcommand{\dd}{{\rm d}}
\newcommand{\lin}{{\rm lin}}
\newcommand{\low}{{\rm low}}

\begin{document}
\title{The cluster-galaxy cross-spectrum:}
\subtitle{an additional probe of cosmological and halo parameters}
\author{Gert H\"utsi \inst{1,2} \and Ofer Lahav\inst{1}}
\institute{Department of Physics and Astronomy, University College London, London, WC1E 6BT\\
  \email{ghutsi@star.ucl.ac.uk, lahav@star.ucl.ac.uk} \and Tartu Observatory, T\~oravere 61602, Estonia}
\date{Received / Accepted}
\titlerunning{The cluster-galaxy cross-spectrum}

\abstract
{There are several wide field galaxy and cluster surveys planned for the nearest future, e.g. BOSS, WFMOS, ADEPT, Hetdex, SPT, eROSITA. In the simplest approach one would analyze these independently, thus neglecting the extra information provided by the cluster-galaxy cross-pairs.}
{In this paper we have focused on the possible synergy between these surveys by investigating the amount of information encoded in the cross-pairs.}
{We present a model for the cluster-galaxy cross-spectrum within the Halo Model framework. To assess the gain in performance due to inclusion of the cluster-galaxy cross-pairs we carry out a Fisher matrix analysis for a BOSS-like galaxy redshift survey targeting luminous red galaxies and a hypothetical mass-limited cluster redshift survey with a lower mass threshold of $1.7\times10^{14}\,h^{-1}M_{\odot}$ over the same volume.}
{On small scales cluster-galaxy cross-spectrum probes directly density profile of the halos, instead of the density profile convolved with itself, as is the case for the galaxy power spectrum. Due to this different behavior, adding information from the cross-pairs helps to tighten constraints on the halo occupation distribution (e.g. a factor of $\sim 2$ compression of the error ellipses on the $m_g^{\low}$-$\alpha$ plane) and offers an alternative mechanism compared with techniques that directly fit halo density profiles.
By inclusion of the cross-pairs a factor of $\sim 2$ stronger constraints are obtained for $\sigma_8$, while the improvement for the dark energy figure-of-merit is somewhat weaker: an increase by a factor of $1.4$.

We have also written down the formalism for the case when only photometric redshifts are available for both the clusters and the galaxies. For the analysis of the photometric surveys the inclusion of the cluster-galaxy cross-pairs might be very beneficial since the photo-z errors for the clusters are usually significantly smaller than that for the typical galaxies.}
{}
\maketitle

\section{Introduction}
With the advent of the modern wide field spectroscopic sky surveys, which have measured redshifts for hundreds of thousands of galaxies, e.g. 2dFGRS\footnote{http://www.mso.anu.edu.au/2dFGRS/}, SDSS\footnote{http://www.sdss.org/}, we have gained a good picture of the cosmic large-scale structure out to redshifts of $\sim 0.5$. From the measurements of the cosmic microwave background angular temperature fluctuations (e.g. \citealt{2008arXiv0803.0732H}) we have obtained a good understanding of the nature and statistics of the small density fluctuations that through gravitational instability have evolved to the filament-void network of the matter surrounding us at low redshifts. These measurements have proven that initial fluctuations are compatible of being adiabatic and Gaussian with roughly the scale-free initial spectrum. As the initial fluctuations can be described as a Gaussian random field the full information is contained in the two-point correlator of the field. Under the linear evolution of the initial fluctuation field, which is valid on large scales ($k \lesssim 0.1\, h\,\rm{Mpc}^{-1}$), the Gaussian field remains Gaussian in nature, and as such explains the central interest in measuring the clustering power spectrum (or its real space analog -- two-point correlation function) using the high quality data from the above-mentioned spectroscopic surveys. Data from the 2dFGRS and from the SDSS survey have both lead to the high precision measurement of the low-redshift galaxy power spectra \citep{2001MNRAS.327.1297P,2004ApJ...606..702T,2005MNRAS.362..505C,2006PhRvD..74l3507T}. Moreover, these datasets have proven to be powerful enough to enable one to detect theoretically expected small fluctuations ($\sim 5\%$ relative amplitude) -- baryonic acoustic oscillations (BAO) -- in the clustering power spectra \citep{2005ApJ...633..560E,2005MNRAS.362..505C,2006A&A...449..891H,2007MNRAS.378..852P,2007MNRAS.374.1527B,2006PhRvD..74l3507T,2007ApJ...657..645P}. The realization that BAO can be used as a standard ruler to map the low-redshift expansion history of the Universe \citep{1998ApJ...496..605E,2003ApJ...594..665B,2003PhRvD..68f3004H,2003PhRvD..68h3504L,2003ApJ...598..720S}, and thus to shed light on the nature of the mysterious dark energy (DE), has lead several groups to propose next generation redshift surveys that should cover even larger sky areas and go significantly deeper in redshift. A few examples of the proposed projects include BOSS\footnote{http://cosmology.lbl.gov/BOSS/}, WFMOS\footnote{http://arxiv.org/ftp/astro-ph/papers/0510/0510272.pdf}, ADEPT\footnote{e.g. http://www.science.doe.gov/hep/hepap/feb2007/\\hepap\_bennett\_feb07.pdf}, Hetdex\footnote{http://www.as.utexas.edu/hetdex/}, and also already ongoing WiggleZ\footnote{http://wigglez.swin.edu.au/} project performed on Anglo-Australian telescope. 

In addition to these galaxy redshift surveys a fast development in the microwave detector technology has lead to the possibility of detecting tens of thousands of galaxy clusters over several thousand square degrees through thermal Sunyaev-Zeldovich (SZ) effect \citep{1972CoASP...4..173S,1980ARA&A..18..537S}. Currently the most ambitious amongst these projects is the ongoing SPT\footnote{http://pole.uchicago.edu/} survey. Also there are plans in the X-ray community to carry out next generation X-ray cluster survey with the yield of $\sim 100,000$ galaxy clusters (eROSITA\footnote{http://www.mpe.mpg.de/erosita/MDD-6.pdf} survey). With a further possibility of obtaining spectroscopic redshifts for the SZ/X-ray selected galaxy clusters through the optical follow-up, one can turn these large cluster samples to an extremely powerful cosmological probe.

In this paper we study the possible synergy between these galaxy and galaxy cluster surveys, particularly focusing on the two-point clustering measures. As the simplest approach one would analyze these surveys independently, i.e. concentrating separately on galaxy-galaxy and cluster-cluster pairs. However, this approach would neglect extra information provided by the cluster-galaxy cross-pairs. In fact, the study of the cluster-galaxy cross-correlations has rather long history, e.g. \citealt{1974ApJS...28...37P,1977ApJ...214L...1S,1977ApJ...215..703S,1988MNRAS.231..635L}. Although rather sparse, cluster samples with strongly increased clustering strength as compared to the galaxies, had proven to be powerful tracers of the large-scale structure of the Universe. Initially the main interest in determining the cluster-galaxy cross-correlation function was driven by the desire to reduce the shot noise level inherent to these relatively sparse cluster samples.

As the main task of this work we are going to determine the information gain provided by the cluster-galaxy cross-spectrum ($P_{c-g}$) over the information encoded in the galaxy-galaxy and cluster-cluster spectra ($P_{g-g}$ and $P_{c-c}$, respectively). We carry out our modeling using a Halo Model (HM) (e.g. \citealt{2000MNRAS.318..203S,2001MNRAS.325.1359S,2002PhR...372....1C}) approach. To our knowledge this is the first time when HM has been used to model $P_{c-g}$. As such, we try to give a rather comprehensive explanation of the results obtained. We apply the derived formalism to a BOSS-like galaxy survey covering $10,000$ deg$^2$ and reaching redshifts of $0.7$. For the cluster survey we assume that it covers the same volume as the galaxy survey and is simply mass-limited with the lower mass cut-off $m_c^{\low}=1.7\times10^{14}\,h^{-1}M_{\odot}$, which is close to the mass limit obtainable by the SPT. In reality BOSS and SPT surveys do not overlap, but eROSITA, which has roughly the same effective sensitivity (although the threshold mass in this case depends more strongly on redshift) is a full sky cluster survey, and thus has a complete overlap with BOSS.

In our study we focus on two-point clustering statistics. Thus our approach is complementary to the number count analysis. In case of the galaxy cluster samples the complementarity of the number count and clustering analysis has been investigated by \citet{2004ApJ...613...41M}.

Our paper is organized as follows: In Section 2 we present the formalism, which is applied to hypothetical galaxy and cluster cluster surveys in Section 3. There the main focus is to find out the gain in information once cluster-galaxy cross-pairs are included. Finally, Section 4 brings our conclusions.  

Throughout the paper we assume a fiducial cosmology to be a flat ``concordance'' $\Lambda$CDM model \citep{1999Sci...284.1481B} with $\Omega_m=0.27$, $\Omega_b=0.045$, $h=0.7$, and $\sigma_8=0.85$. All these parameter values, except somewhat larger value for $\sigma_8$, are in good agreement with the recent WMAP5 results \citep{2008arXiv0803.0586D}.  

\section{Halo Model power spectra. Covariances and Fisher matrices}\label{sec2}
\subsection{Power spectra in real space}\label{sec2.1}
In this Subsection we present the results for the matter ($m$-$m$), galaxy ($g$-$g$), cluster ($c$-$c$), and cluster-galaxy ($c$-$g$) power spectra in real space within the framework of the Halo Model (HM). The corresponding results for the redshift-space are given in the following Subsection. For a comprehensive review on HM see e.g. \citet{2002PhR...372....1C}. To the readers familiar with the HM we suggest to directly jump to Eqs. (\ref{eq5})-(\ref{eq8}), which present a commonly used formulation for the $g-g$ power spectrum in real space. Our formalism for the $c-c$ and $c-g$ spectra in real space is new and given by Eqs. (\ref{eq13})-(\ref{eq18}). Eqs. (\ref{eq23})-(\ref{eq34}) provide analogous formulae for the anisotropic spectra in redshift-space.

\begin{figure*}
\centering
\includegraphics[width=\plotwdtwo]{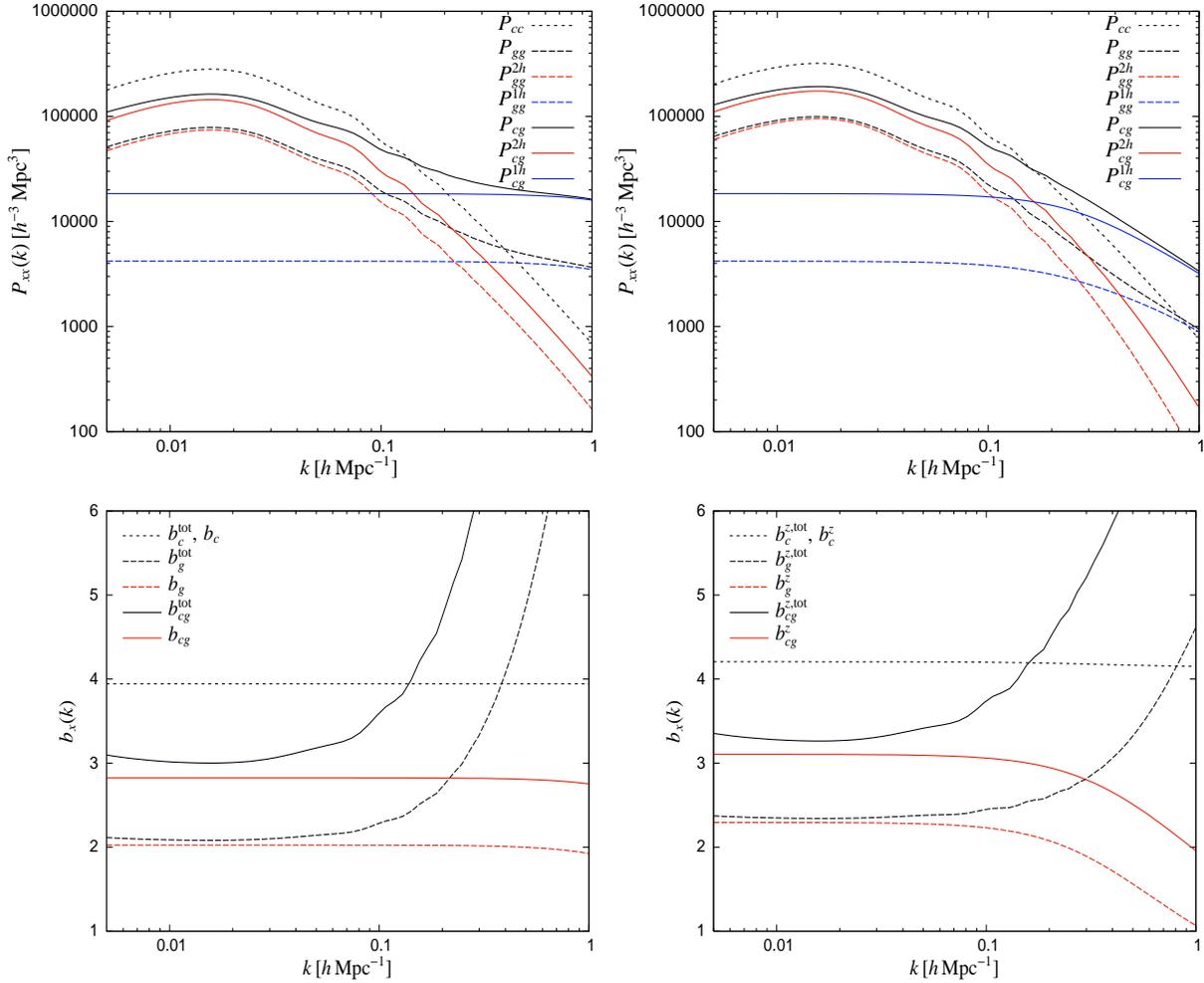}
\caption{Power spectra and bias parameters in real- (left-hand panels) and redshift-space (right-hand panels). The short-dashed, dashed, and solid lines correspond to cluster, galaxy, and cross-spectrum, respectively.  Where relevant we have shown contributions from the $1h$ and $2h$ terms separately. In lower panels a superscript ``tot'' is used to denote a total bias parameter, which includes both $1h$ and $2h$ contributions, in contrast to the bias parameters $b_g$, $b_c$, and $b_{cg}=\sqrt{b_g b_c}$, which incorporate $2h$ contributions only. See the main text for the assumed values of the HOD and other parameters.}
\label{fig01}
\end{figure*}

In HM the power spectra can be expressed as a sum of the one-halo ($1h$) and two-halo ($2h$) terms:

\begin{equation}  
P_x = P^{1h}_x + P^{2h}_x\,,
\end{equation}
which provide the contributions of pairs of points inhabiting the same or separate dark matter halos, respectively. For our cases of interest the subscript $x$ can be either $m$-$m$, $g$-$g$, $c$-$c$, or $c$-$g$.

Following \citet{2000MNRAS.318..203S} and \citet{2002PhR...372....1C} the $1h$ and $2h$ terms for the matter power spectrum at redshift $z$ can be expressed as:  
\begin{eqnarray}
 P^{1h}_{m-m}(k|z) &=& \int \dd m\,n(m|z) \left(\frac{m}{\bar{\rho}}\right)^2 u_m^2(k|m,z)\,,\\
 P^{2h}_{m-m}(k|z) &=& b^2_m(k|z) g^2(z) P^\lin(k|z=0)\,, \\
b_m(k|z) &\equiv& \int \dd m\, n(m|z) b(m|z) \left(\frac{m}{\bar{\rho}}\right) u_m(k|m,z)\,.
\end{eqnarray}
In the above expressions $n(m|z)$ is the mass function and $b(m|z)$ halo bias parameter at redshift $z$. We calculate these using prescriptions by \citet{1999MNRAS.308..119S} and \citet{2001MNRAS.323....1S}. $g(z)$ represents linear growth factor at redshift $z$ normalized such that $g(z=0)=1$, and $\bar{\rho} \equiv \Omega_m\rho_c=\int \dd m\, m n(m|z) b(m|z)$ gives the mean comoving matter density. $u_m(k|m,z)$ is the normalized (i.e. $\lim \limits_{k\rightarrow 0} u_m(k|m,z) = 1$) Fourier transform of the dark matter density profile within a halo of mass $m$ at redshift $z$, which we take to be given by the the NFW \citep{1997ApJ...490..493N} form. The concentration parameter of the density profiles and its evolution with redshift is assumed to take an analytic form as given in \citet{2001MNRAS.321..559B}. To calculate linear power spectrum $P^\lin(k|z=0)$ we assume power-law initial spectrum with spectral index $n_s=1$ and use fitting formulae for transfer functions $T(k)$ as given by \citet{1998ApJ...496..605E}.

The corresponding results for the $g$-$g$ case can be expressed as (see \citet{2002PhR...372....1C} for details): 
\begin{eqnarray}
 P^{1h}_{g-g}(k|z) &=& \int \dd m\,n(m|z) \frac{\langle N_g(N_g-1)|m \rangle}{\bar{n}^2_g(z)} |u_g(k|m,z)|^p\,,\label{eq5}\\
 P^{2h}_{g-g}(k|z) &=& b^2_g(k|z) g^2(z) P^\lin(k|z=0)\,, \label{eq6}\\
b_g(k|z) &\equiv& \int \dd m\, n(m|z) b(m|z) \frac{\langle N_g|m \rangle}{\bar{n}_g(z)} u_g(k|m,z)\,, \label{eq7}\\
\bar{n}_g(z) &\equiv& \int \dd m\, n(m|z) \langle N_g|m \rangle\,. \label{eq8}
\end{eqnarray}
These results follow directly from the corresponding results for the $m$-$m$ case if one imagines the mass density field as composed of discrete particles, and thus can write $m/\bar{\rho}\equiv N/\bar{n}$ where $N$ is the number of particles in a halo with mass $m$ and $\bar{n}$ is the average number density of particles. To count the number of pairs correctly in this discrete case we have a second factorial moment (instead of simply a second moment) of the Halo Occupation Distribution (HOD) in Eq. (\ref{eq5}). In the above relations $\bar{n}_g(z)$ is the mean number density of galaxies and $u_g(k|m,z)$ is the normalized Fourier transform of the galaxy number density profile within a halo of mass $m$ at redshift $z$. For $u_g(k|m,z)$ we assume that it follows the corresponding density distribution for the dark matter i.e. we take $u_g(k|m,z)=u_m(k|m,z)$.
We further assume that the 1st and 2nd factorial moments of the HOD are given by the following parametric forms (e.g. \citealt{2004MNRAS.348..250C}):
\begin{eqnarray}
\langle N_g|m \rangle &=& \left \{
\begin{array}{cc}
\left ( \frac{m}{m_0} \right )^\alpha & {\rm\quad if \quad} m\geq m_g^{\low} \\
0 & {\rm\quad if \quad} m < m_g^{\low}\,,\label{eq9}
\end{array}
\right. \\
\langle N_g(N_g-1)|m \rangle & = & \beta^2(m) \langle N_g|m \rangle ^2\,,\label{eq10}\\
\beta(m) & = & \left \{ 
\begin{array}{ll}
\frac{1}{2} \log \left ( \frac{m}{10^{11} \ h^{-1}M_{\odot}} \right ) & \mathrm{\, if \,} m<10^{13} \  h^{-1}M_{\odot} \\
1 & \mathrm{\, otherwise\,.}\label{eq11}
\end{array}
\right.
\end{eqnarray}
This HOD model is consistent of assuming that for the high mass halos ($m>10^{13} \  h^{-1}M_{\odot}$) the HOD follows Poisson distribution whereas for the smaller masses the fluctuations in $N_g$ are sub-Poissonian. Thus in our case we have three free parameters describing the HOD: $\alpha$, $m_0$, and $m_g^{\low}$.
 
Here we have made an assumption that if there are any galaxies in a halo then one of these is located in the central position. Thus for the $1h$ term we have to deal with two types of pairs: (i) ones that involve the central object, and so probe directly the density profile of the halo (i.e. the power law index $p=1$ in Eq. (\ref{eq5})), (ii) all the other pairs, which describe the convolution of the density profile with itself (i.e. $p=2$). The detailed treatment (see e.g. \citealt{2002PhR...372....1C}) of the central objects leads to the following choice for $p$ in Eq. (\ref{eq5}):
\begin{equation}
p  = \left\{ 
\begin{array}{ll}
1 & \mathrm{\quad if \quad} \langle N_g(N_g-1)|m \rangle < 1\\
2 & \mathrm{\quad if \quad} \langle N_g(N_g-1)|m \rangle \geq 1\,.
\end{array}
\right.
\end{equation}
   
Following the derivations for the matter and galaxy power spectra as given in \citet{2002PhR...372....1C} we obtain the corresponding results for the $c$-$c$ and $c$-$g$ cases. The cluster power spectrum can be given as:
\begin{eqnarray}
 P^{1h}_{c-c}(k|z) &=& 0\,, \label{eq13}\\
 P^{2h}_{c-c}(k|z) &=& b^2_c(k|z) g^2(z) P^\lin(k|z=0)\,, \\
b_c(k|z) &\equiv& \frac{1}{\bar{n}_c(z)}\int \limits_{m_c^{\low}(z)} \dd m\, n(m|z) b(m|z)\,, \label{eq15}\\
\bar{n}_c(z) &\equiv& \int \limits_{m_c^{\low}(z)} \dd m\, n(m|z)\,,
\end{eqnarray}
and the cluster-galaxy cross-spectrum as:
\begin{eqnarray}
P^{1h}_{c-g}(k|z) &=& \frac{1}{\bar{n}_c(z) \bar{n}_g(z)} \int \limits_{m_c^{\low}(z)} \dd m\,n(m|z) \left< N_g|m\right> u_g(k|m,z)\,, \\
P^{2h}_{c-g}(k|z) &=& b_c(k|z) b_g(k|z) g^2(z) P^\lin(k|z=0)\,. \label{eq18} 
\end{eqnarray}
It is clear that there is no $1h$ term for the cluster power spectrum, \footnote{Here we have chosen to subtract the shot noise term, which effectively corresponds to the cluster number counts. Thus our clustering analysis is independent and complementary to the number count studies.} whereas the $2h$ term can be obtained from the corresponding $g$-$g$ $2h$ term by replacing $\langle N_g|m \rangle$ with $1$ (i.e. each dark matter halo hosts one cluster), $\bar{n}_g(z)$ with $\bar{n}_c(z)$, and taking the normalized Fourier transform of the density profile ($\delta$-function in our case) equal to $1$. \footnote{To be more precise, even in Eq. (\ref{eq7}) one has to take $u_g(k|m,z)=1$ if $\langle N_g|m \rangle \leq 1$, since as we assumed, if there is one galaxy per halo, it must be located at the central position. In reality, however, these details make a negligible difference.} For calculating cluster power spectra we assume for simplicity that our cluster sample is mass-selected above some mass limit $m_c^{\low}$ which can be dependent on redshift.   

On small scales (i.e. relevant for the $1h$ term) cluster-galaxy cross correlations probe density profile around halo center i.e. in comparison to the $g-g$ case, where power law index $p$ in Eq. (\ref{eq5}) can take value $2$, here we have only a linear dependence on $u_g(k|m,z)$. The number of pairs contributing to the $1h$ term is simply given by the number of galaxies within each halo i.e. $\left< N_g|m\right>$.

In the above formulae we have used the notations $b_m(k|z)$, $b_g(k|z)$, and $b_c(k|z)$ to emphasize the fact that these functions give the effective bias parameters for large scales (i.e. neglecting the contributions from $1h$ terms). From Eq. (\ref{eq18}) we see that the effective large-scale bias parameter $b_{c-g}(k|z)$ for the cross-spectrum is given by $\sqrt{b_c(k|z)b_g(k|z)}$, as expected.

Examples of the real-space spectra and bias parameters are given on the left-hand panels of Fig. \ref{fig01}. The results presented there correspond to the redshift $z=0.5$. To calculate the cluster spectrum we have assumed a mass-limited sample of clusters above a threshold mass $m_c^{\low}=1.7\times10^{14}\,h^{-1}M_{\odot}$. The parameters for the galaxy HOD were fixed as follows: $\alpha=0.9$, $m_g^{\low} = 5 \times 10^{12}\,h^{-1}M_{\odot}$, and $m_0 = 4.2 \times 10^{13}\,h^{-1}M_{\odot}$. For the motivation of these values see Sec. \ref{sec3}. In Fig. \ref{fig01} the short-dashed, dashed, and solid lines correspond to cluster, galaxy, and cross-spectrum, respectively. Where relevant we have also shown contributions from the $1h$ and $2h$ terms separately. Here it is worth noticing a relatively bigger importance of the $1h$ term in case of the cross-spectrum as compared to the galaxy-galaxy power spectrum. In the lower panel we have used a superscript ``tot'' to denote a total bias parameter, which includes both $1h$ and $2h$ contributions, in contrast to the bias parameters $b_g$, $b_c$, (see Eqs. (\ref{eq7}),(\ref{eq15})) and $b_{cg}=\sqrt{b_g b_c}$, which incorporate $2h$ contributions only. For small values of the wavenumber $k$ the total bias parameters deviate slightly from the results obtained by using $2h$ terms only due to the large-scale ``leakage'' of the $1h$ term.\footnote{It is clear that these additional Poissonian fluctuations on the largest scales are incompatible with the large-scale homogeneity requirement. This is a shortcoming of the HM that hasn't been solved satisfactorily so far. A possible way around it would be the use of compensated density profiles for the halos, i.e. no DC component in their Fourier transforms. Even then, in $2h$ terms one still has to use uncompensated profiles, since convolving with the profile with zero DC component would damp out the large-scale power completely. However, for the scales of interest in this paper the deviations of the bias parameters are rather small, and thus we do not make an attempt to correct for this shortcoming.}   

\subsection{Redshift-space power spectra}
In this Subsection we are going to generalize the previous results to include the redshift-space distortions. Since in addition to the above-mentioned future redshift surveys there are several imaging surveys planned (e.g. DES\footnote{https://www.darkenergysurvey.org/}, Pan-STARRS\footnote{http://pan-starrs.ifa.hawaii.edu/public/}, DUNE\footnote{http://www.dune-mission.net/}, LSST\footnote{http://www.lsst.org/}), in Appendix \ref{appb} we also give a simple treatment in case only photometric redshifts are available. Here we partially follow the formalism as given in \citet{2001MNRAS.321....1W} and in \citet{2001MNRAS.325.1359S}. For the results presented in this Subsection we assume a flat sky approximation.

Small-scale redshift-space distortions (``Fingers-of-God'' (FOG)) in coordinate space can be modelled as a convolution along the line of sight with a 1D Gaussian kernel. As Gaussian stays Gaussian also in Fourier space the matter density contrast gets modified as:
\begin{equation}
\delta^m_{\mathbf{k}} \rightarrow \delta^m_{\mathbf{k}}\cdot \mathcal{F}(k\sigma_m \mu)\,, \label{eq19}   
\end{equation}    
where the damping factor $\mathcal{F}$ is given by:
\begin{equation}
\mathcal{F}(x) \equiv \exp\left(-\frac{x^2}{2}\right)\,.
\end{equation}    
Here $\mu$ is the cosine of the angle between the line of sight direction $\hat{\mathbf{r}}$ and wavevector $\mathbf{k}$, i.e. $\mu \equiv \hat{\mathbf{r}}\cdot \hat{\mathbf{k}}$. As in \citet{2004MNRAS.348..250C}, we assume that the one dimensional velocity dispersion of matter inside a halo with mass $m$ at redshift $z$, $\sigma_m(m,z)$, follows the scaling of the isothermal sphere model:
\begin{equation}
\sigma_m(m,z) = \gamma \cdot \sqrt{\frac{Gm}{2r_{{\rm vir}}(m,z)}}\,,
\end{equation}    
where $r_{{\rm vir}}$ is the virial radius of the halo and $\gamma$ is a free parameter, which in the following we keep fixed to $1$, i.e. assume precisely the value for the isothermal sphere model. The corresponding velocity dispersion of galaxies is taken to follow the same relation.     

There are additional redshift-space distortions on large scales due to the coherent inflows of matter towards massive accretion centers. This leads to the modification of the density fluctuations in Fourier space such that \citep{1987MNRAS.227....1K}:
\begin{equation}
\delta^x_{\mathbf{k}} \rightarrow \delta^x_{\mathbf{k}} + f\cdot \delta^m_{\mathbf{k}}\mu^2\,, 
\end{equation}
where $f$ represents the logarithmic derivative of the growth factor $g(z)$, i.e. $f\equiv \dd\log g/\dd\log a$, and superscript $x$ can be either $m$ (for matter), $c$ (clusters), or $g$ (galaxies).

Taking into account both the small and large-scale distortions we can write down the results for the anisotropic power spectra as follows: 

\begin{eqnarray}
 P^{1h}_{c-c}(k,\mu|z) &=& 0\,, \label{eq23}\\
 P^{1h}_{g-g}(k,\mu|z) &=& \int \dd m\,n(m|z) \frac{\langle N_g(N_g-1)|m \rangle}{\bar{n}^2_g(z)} |u_g(k|m,z)|^p \nonumber \\ &\cdot& \mathcal{F}^p\left[k\sigma_m(m,z)\mu\right]\,,\label{eq24}\\
 P^{1h}_{c-g}(k,\mu|z) &=& \frac{1}{\bar{n}_c(z) \bar{n}_g(z)} \int \limits_{m_c^{\low}(z)} \dd m\,n(m|z) \left< N_g|m\right> u_g(k|m,z) \nonumber \\ &\cdot& \mathcal{F}\left[k\sigma_m(m,z)\mu\right]\,,\label{eq25}\\
 P^{2h}_{c-c}(k,\mu|z) &=& b_c^z(k,\mu|z)^2 g^2(z) P^\lin(k|z=0)\,,\\
 P^{2h}_{g-g}(k,\mu|z) &=& b_g^z(k,\mu|z)^2 g^2(z) P^\lin(k|z=0)\,,\\
 P^{2h}_{c-g}(k,\mu|z) &=& b_{cg}^z(k,\mu|z)^2 g^2(z) P^\lin(k|z=0)\,,\\
 b_c^z(k,\mu|z) &\equiv& b_c(k,\mu|z) + f(z) b_m(k,\mu|z)\mu^2\,\label{eq29}\\
 b_g^z(k,\mu|z) &\equiv& b_g(k,\mu|z) + f(z) b_m(k,\mu|z)\mu^2\,\label{eq30}\\
 b_{cg}^z(k,\mu|z) &\equiv& \sqrt{b_c^z(k,\mu|z)b_g^z(k,\mu|z)}\,\label{eq31}\\
 b_c(k,\mu|z) &\equiv& \frac{1}{\bar{n}_c(z)}\int \limits_{m_c^{\low}(z)} \dd m\, n(m|z) b(m|z)\,,\label{eq32}\\
b_g(k,\mu|z) &\equiv& \int \dd m\, n(m|z) b(m|z) \frac{\langle N_g|m \rangle}{\bar{n}_g(z)} u_g(k|m,z) \nonumber \\ &\cdot& \mathcal{F}\left[k\sigma_m(m,z)\mu\right]\,,\label{eq33}\\
b_m(k,\mu|z) &\equiv& \int \dd m\, n(m|z) b(m|z) \left(\frac{m}{\bar{\rho}}\right) u_m(k|m,z) \nonumber \\ &\cdot& \mathcal{F}\left[k\sigma_m(m,z)\mu\right]\,.\label{eq34}
\end{eqnarray}

Examples of the spectroscopic redshift-space spectra and bias parameters are given on the right-hand panels of Fig. \ref{fig01}. The HOD and other parameters are exactly the same as on the left-hand panels (see the end of Subsec. \ref{sec2.1} for details), which show the corresponding results for the real-space. As before, a superscript ``tot'' on lower panel denotes a total bias parameter, which includes both $1h$ and $2h$ contributions, in contrast to the bias parameters $b_c^z$, $b_g^z$, and $b_{cg}^z$ (see Eqs. (\ref{eq29}),(\ref{eq30}),(\ref{eq31})), which incorporate $2h$ contributions only. The spectra and bias parameters presented here are the angle-averaged versions. Note the expected large-scale boost and small-scale damping of power as compared to the equivalent real-space results.  

\subsection{Cosmological distortion}
The cosmological distortion \citep{1979Natur.281..358A} arises due to the simple fact that conversion of the observed redshifts to comoving distances requires the specification of the cosmological model. If this cosmology differs from the true one, we are left with additional distortion of distances along and perpendicular to the line of sight. In general, the spatial power spectrum measurements, in contrast to the angular spectra, are model dependent i.e. along with the measurements of the 3D power spectrum one always has to specify the so-called fiducial model used to analyze the data. In principle, for each of the assumed cosmological models one should redo the full power spectrum analysis to accommodate different distance-redshift relation. However, there is an easier way around: one can find an approximate analytical transformation that describes how the model spectrum should look like under the distance-redshift relation given by the fiducial model. This transformation for the power spectrum is different along and perpendicular to the line of sight, and also is dependent on redshift. For more discussion on these issues see e.g. \citet{2006A&A...459..375H}. The observed spectrum $\widetilde{P}_x$ under the assumed distance-redshift relation of the fiducial cosmological model is related to the true spectrum $P_x$ as follows \citep{2006A&A...459..375H}:
\begin{equation}  
\widetilde{P}_x(k_{\parallel},k_{\perp}|z)=\frac{1}{c_{\parallel}(z)\cdot c_{\perp}^2(z)}\cdot P_x\left[\frac{k_{\parallel}}{c_{\parallel}(z)},\frac{k_{\perp}}{c_{\perp}(z)}|z\right]\,,
\end{equation}  
where the distortion parameters along and perpendicular to the line of sight are given as:
\begin{eqnarray}
c_{\parallel}(z)&=&\frac{H^{\mathrm{fid}}(z)}{H(z)}\,,\\
c_{\perp}(z)&=&\frac{d_{\perp}(z)}{d_{\perp}^{\mathrm{fid}}(z)}\,.
\end{eqnarray}
Here $H(z)$ denotes the Hubble parameter and $d_{\perp}(z)$ is the comoving angular diameter distance. Superscript $^{\mathrm{fid}}$ refers to the fiducial model. Here and in the following we use a tilde on top of $P_x$ to denote theoretical spectrum ``transformed to the reference frame of the fiducial cosmology''. As we use the spectra that have the dimensions of volume an extra division by $c_{\parallel}(z)\cdot c_{\perp}^2(z)$ occurs due to the transformation of the volume elements. The same transformation in terms of $k=(k_{\parallel}^2+k_{\perp}^2)^{1/2}$ and $\mu=k_{\parallel}/k$ can be expressed as:
\begin{equation}  
\widetilde{P}_x(k,\mu|z)=\frac{1}{c_{\parallel}(z)\cdot c_{\perp}^2(z)}\cdot P_x \left[ \frac{\alpha(\mu,z)}{c_{\perp}(z)}k,\frac{1}{\varkappa(z)\alpha(\mu,z)}\mu|z\right]\,, 
\end{equation}  
where  
\begin{eqnarray}
\varkappa(z) &\equiv& \frac{c_{\parallel}(z)}{c_{\perp}(z)}\,,\\
\alpha(\mu,z) &\equiv& \sqrt{1+\left( \frac{1}{\varkappa^2(z)} -1 \right)\mu^2}\,.
\end{eqnarray}

\subsection{Covariances and Fisher matrices}
Having a HM description for the various spectra we can go on to calculate covariances and Fisher matrices. We give all the necessary details in Appendix \ref{appa}.   
 
\begin{figure}
\centering
\includegraphics[width=\plotwd]{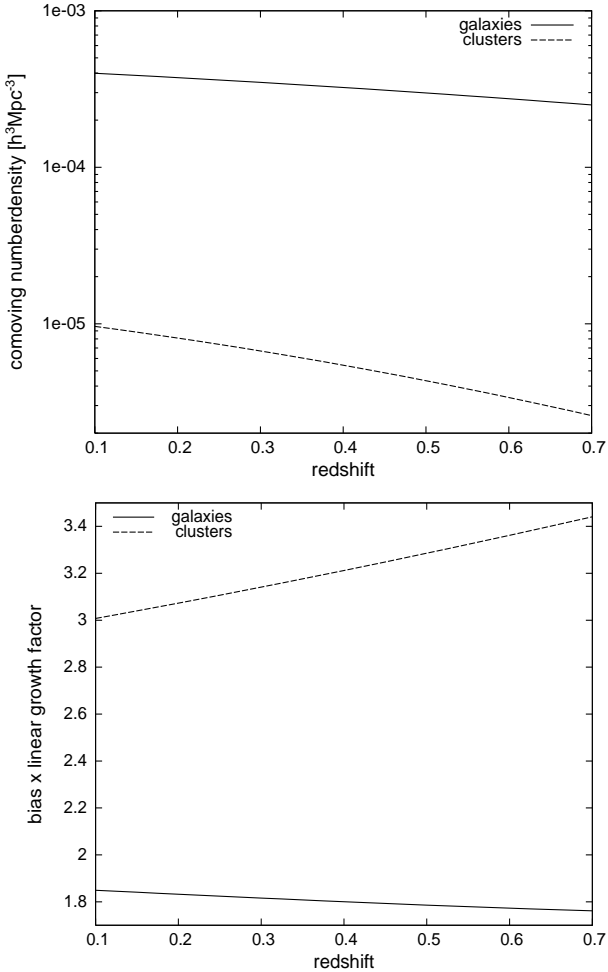}
\caption{Comoving number density (upper panel) and the product of the real-space bias parameter and linear growth factor $b(z)g(z)$ (lower panel) for galaxies (solid lines) and galaxy clusters (dashed lines) as a function of redshift.}
\label{fig02}
\end{figure}

\begin{figure}
\centering
\includegraphics[width=\plotwd]{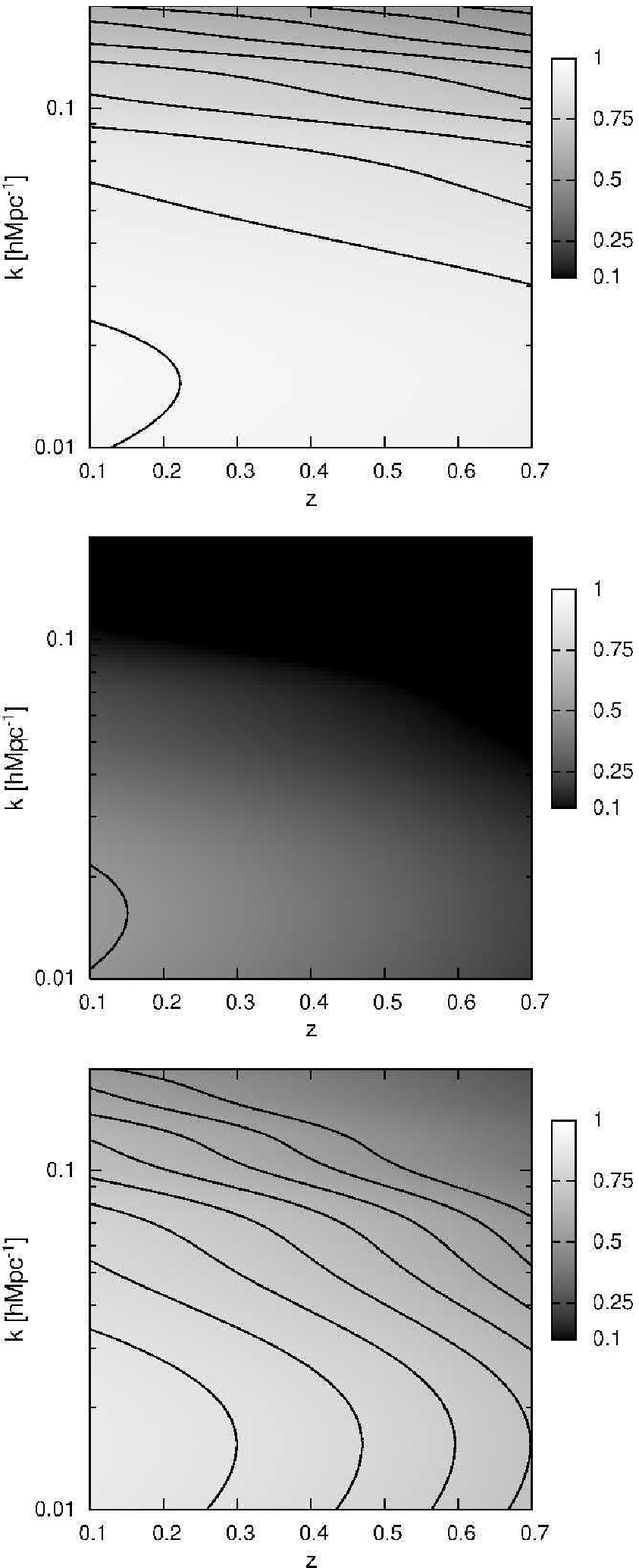}
\caption{The angle-averaged weight factors $w^g$ (upper panel), $w^c$ (middle panel), and $w^{c-g}$ (lower panel) as functions of the comoving wavenumber $k$ and redshift $z$. The discrete contour lines start from the value of $0.5$ and increase with a constant step of $0.05$.} 
\label{fig03}
\end{figure}

\section{Application of the formalism}\label{sec3}
In this Section we are going to apply the machinery presented in the previous section to the hypothetical ideal galaxy and galaxy cluster redshift surveys. For the galaxy survey we assume specifications similar to the planned BOSS survey: $10,000$ deg$^2$ sky coverage, redshift range $z=0.1-0.7$, and the comoving number density of galaxies $\sim 3\times 10^{-4} \,h^3\,\rm{Mpc}^{-3}$. Thus the volume of the survey is $\sim 5.7 \,h^{-3}\,\rm{Gpc}^{3}$ with the total number of galaxies reaching $\sim 1,700,000$. As mentioned in Sec. \ref{sec2} we assume a simple HOD given by Eq. (\ref{eq9}), i.e. we have three free parameters: $\alpha$, $m_0$, and $m_g^{\low}$. For the parameter $\alpha$ we choose a fiducial value of $0.9$, which is typical for the red galaxies according to the semi-analytic galaxy formation models (see e.g. \citealt{2002PhR...372....1C})\footnote{However, from the observational data several authors have obtained values for $\alpha$ which are somewhat higher: \citet{2005MNRAS.361..415C} found $\alpha=1.05$ for the 2dFGRS red galaxies, \citet{2008MNRAS.385.1257B} found $\alpha$ for the LRGs to be evolving with redshift, from $1.57$  to $1.80$.}. Looking at the equations given in the previous section it is evident that power spectra do not depend on $m_0$. Thus after fixing the value for $\alpha$ the fiducial value for the parameter $m_g^{\low}$ can be chosen such as to give a clustering strength typical of the luminous red galaxies targeted by BOSS, i.e. the product of the bias and linear growth factor $b(z)g(z) \sim 2$. This gives us $m_g^{\low} \simeq 5 \times 10^{12}\,h^{-1}M_{\odot}$. Once $\alpha$ and $m_g^{\low}$ are fixed the fiducial value for $m_0$ is obtained through the number density constraint, giving $m_0 \simeq 4.2 \times 10^{13}\,h^{-1}M_{\odot}$. The total survey volume is divided into six redshift bins with a width $\Delta z = 0.1$.  

\begin{figure*}
\centering
\includegraphics[width=\plotwdtwo]{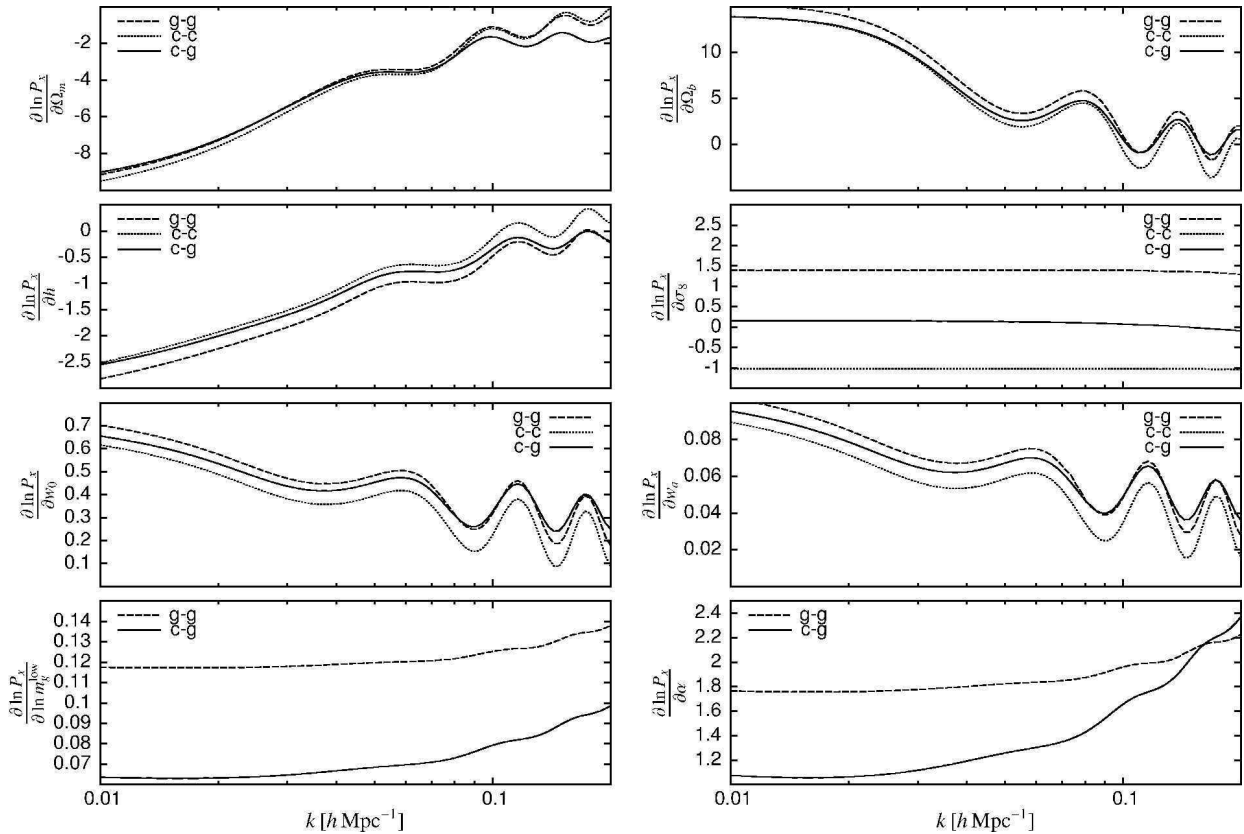}
\caption{Logarithmic derivatives of the galaxy (dashed lines), cluster (dotted lines), and cluster-galaxy power spectra.}
\label{fig04}
\end{figure*}

The galaxy cluster survey is assumed to cover exactly the same volume. For simplicity we model our cluster survey as a mass limited survey above a threshold mass of $1.7\times10^{14}\,h^{-1}M_{\odot}$. This is very similar to the mass sensitivity obtainable by the SPT survey. Of course one could model the cluster selection (e.g. SZ or X-ray selection) more realistically taking the threshold mass to be an appropriate function of redshift. However, for clarity we choose to keep things as simple as possible. As an additional prior we assume that the threshold mass can be calibrated with an accuracy of $10\%$. It turns out that the information in the cluster-galaxy cross-spectrum itself helps in determining the threshold mass with an accuracy of $\sim 4\%$, and thus our results are rather insensitive to this assumed prior.

In addition to the HM parameters we have to fix a fiducial cosmological model. This we take to be a flat $\Lambda$CDM model with $\Omega_m=0.27$, $\Omega_b=0.045$, $h=0.7$, and $\sigma_8=0.85$. We will focus only on spatially flat models in our Fisher matrix calculations. For the dark energy equation of state we use a common $w_0-w_a$ parametrization: $w(a)=w_0+(1-a)w_a$ ($a=1/(1+z)$). So our parameter vector $\mathbf{\Theta}$ has nine components in total: $\mathbf{\Theta}=(\Omega_m,\Omega_b,h,\sigma_8,w_0,w_a,\ln m_g^{\low},\alpha,\ln m_c^{\low})$ with the following fiducial values: $\mathbf{\Theta}^{{\rm fid}}=\left(0.27,0.045,0.7,0.85,-1.,0.,\ln(5 \times 10^{12}),0.9,\ln(1.7 \times 10^{14})\right)$. 

Having fixed the parameters to those fiducial values we end up with a total of $\sim 1,700,000$ galaxies and $\sim 25,000$ galaxy clusters. The redshift dependence of the comoving number density and the product of the real-space linear bias parameter $b(z)$ and the linear growth factor $g(z)$ (i.e. the linear bias parameter with respect to the $z=0$ matter power spectrum) is shown on the upper and lower panels of Fig. \ref{fig02}, respectively. In redshift-space the product $b(z)g(z)$ gets boosted by a factor of $1.12$ and reaches $\sim 2.1$ at redshift $z=0.1$ as compared to the corresponding real-space value of $1.85$.

The important ingredients of the Fisher matrix calculations are the weight factors ($w^g$, $w^c$, $w^{c-g}$) and the logarithmic derivatives of the power spectra (see Eqs. (\ref{eq42}),(\ref{eq43}),(\ref{eq44}),(\ref{eq45})). The angle-averaged weight factors as functions of the comoving wavenumber $k$ and redshift $z$ for the galaxies, clusters, and cluster-galaxy cross-pairs are shown as upper, middle, and lower panel of Fig. \ref{fig03}, respectively. There the discrete contour lines start from the value of $0.5$ and increase with a constant step of $0.05$. As expected, we see that the number density of galaxies ($\sim 3\times 10^{-4} \,h^3\,\rm{Mpc}^{-3}$) is sufficiently high to ensure a well-sampled density field (i.e. weight factors close to $1$) for all of the wavenumbers and redshifts of interest. This is not the case for the galaxy clusters for which the high shot-noise contribution reduces the weights significantly, reaching the value of $0.5$ at large scales only for the lowest redshifts. For the cluster-galaxy cross-pairs the situation is significantly better: here the weight factor is above the value of $0.5$ for the most of the $k-z$ plane.

\begin{figure*}
\centering
\includegraphics[width=\plotwdtwo]{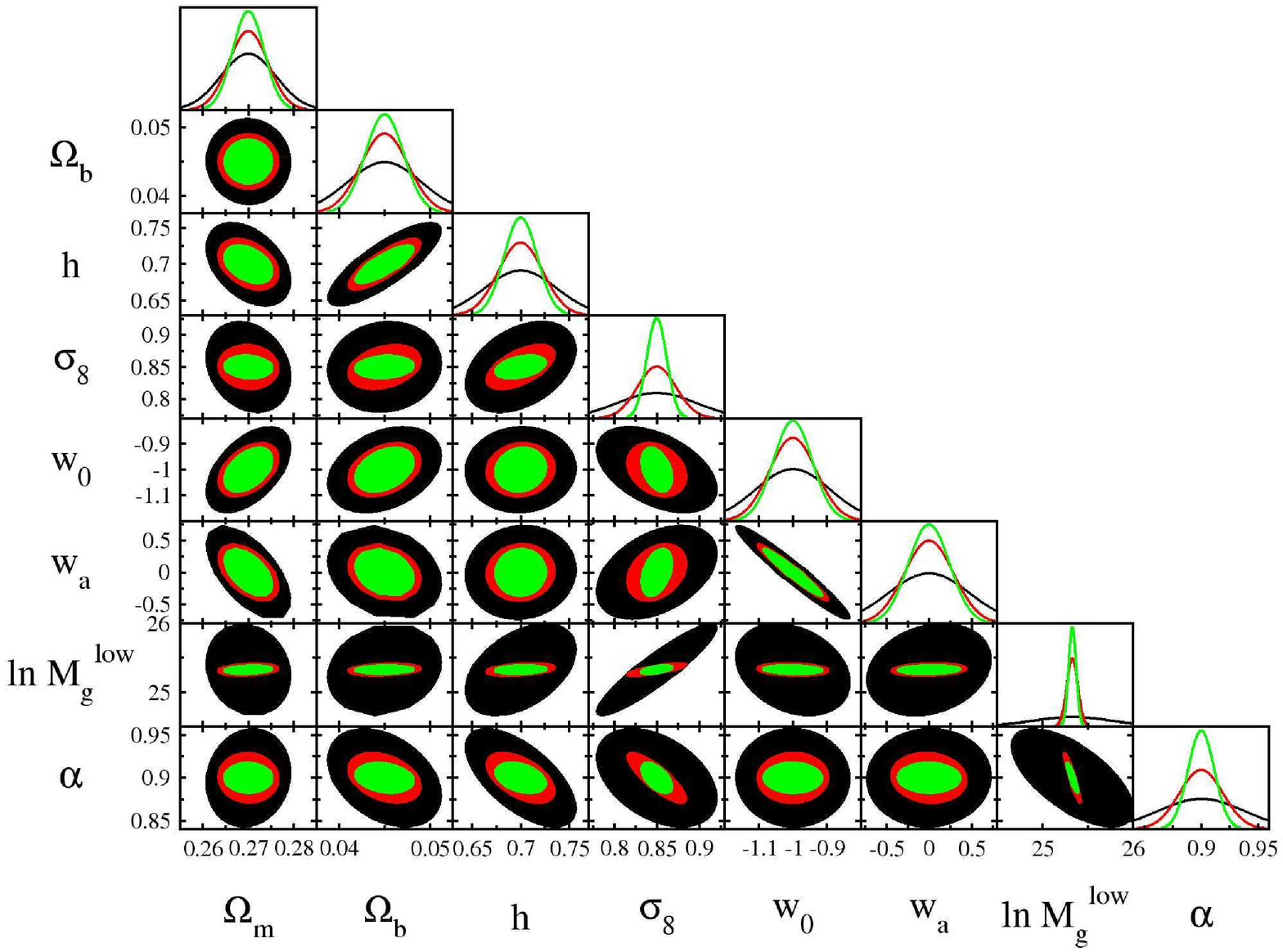}
\caption{2D error ellipses (1-$\sigma$) and 1D Gaussian distributions for eight of the free model parameters. The largest ellipses correspond to the constraints from the cluster-galaxy cross-pairs, middle ones to the galaxy-galaxy pairs, and the smallest to the combined $(g-g)+(c-g)$ case.}
\label{fig05}
\end{figure*}

The logarithmic derivatives of the galaxy, cluster, and cross-spectra are given in Fig. \ref{fig04} as dashed, dotted, and solid lines, respectively. We see that for most of the parameters the derivatives are rather similar, which leads to almost identical degenerate parameter combinations. However, there are significant differences in logarithmic derivatives involving $\sigma_8$\footnote{Note the difference in sign of the logarithmic derivatives of the galaxy and cluster power spectra in this case. By increasing $\sigma_8$ the amplitude of the cluster power spectrum actually drops. This owes to the fact that the higher value of $\sigma_8$ leads to the rapid increase of the number of low mass clusters which are less weakly biased.} and HOD parameters $\alpha$ and $\ln m_g^{\low}$, and thus for these one would expect tighter constraints once the extra information from the cross-pairs is included. This is indeed the case as can be seen from Fig. \ref{fig05} and Table \ref{tab1}. In Fig. \ref{fig05} we have plotted the 2D error ellipses (1-$\sigma$) and also 1D Gaussian distributions for the eight free parameters involved. There the largest ellipses correspond to the constraints from the cross-pairs, middle ones to the galaxy-galaxy pairs, and the smallest to the combined $(g-g)+(c-g)$ case. In Table \ref{tab1} we give the figures-of-merit (FOMs), defined as the inverse areas of the 1-$\sigma$ error ellipses, for several free parameter combinations. The diagonal entries (e.g. $\Omega_m$-$\Omega_m$) in this table give the inverse of the corresponding parameter's 1-$\sigma$ error interval, instead. In each cell the numbers from top to bottom correspond to $c-g$, $g-g$, $(g-g)+(c-g)$, and the ratio of $(g-g)+(c-g)$ and $g-g$ FOMs. In Table \ref{tab1} we have decided not to show an additional gain once $c-c$ information is included, as due to the sparseness of the cluster sample this is relatively modest compared to the accuracy boost brought by the inclusion of the cluster-galaxy cross-pairs. As already mentioned, the biggest boost in accuracy once cross-pairs are included occurs for the parameter combinations involving $\sigma_8$, $\alpha$, and $\ln m_g^{\low}$ as at least one of the parameters, e.g. the error ellipse for the $\sigma_8$-$w_0$ case gets reduced by a factor of $\sim 2.4$, etc. Since the $g-g$ spectrum does not depend on the ninth parameter $\ln m_c^{\low}$ we have decided not to include it in Fig. \ref{fig05} and Table \ref{tab1}. We only note that even without the assumed $10\%$ prior for $m_c^{\low}$ the information encoded in the $c-g$ pairs is able to self-calibrate $m_c^{\low}$ to an accuracy of $\sim 4\%$, and thus our results in Fig. \ref{fig05} and Table \ref{tab1} change only slightly once this extra prior is not applied.  

\begin{figure}
\centering
\includegraphics[width=\plotwd]{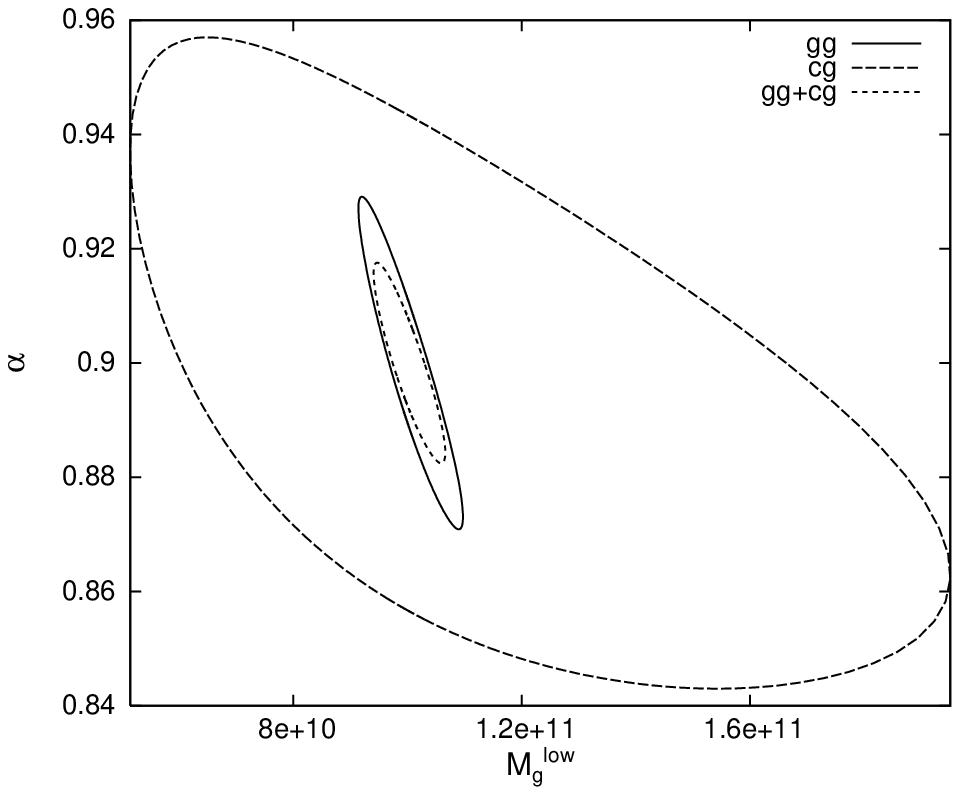}
\caption{1-$\sigma$ error contours on the $m_g^{\low}$-$\alpha$ plane. Solid, dashed, and short-dashed lines correspond to the $g-g$, $c-g$, and $(g-g)+(c-g)$ cases, respectively.}
\label{fig06}
\end{figure}

\begin{figure}
\centering
\includegraphics[width=\plotwd]{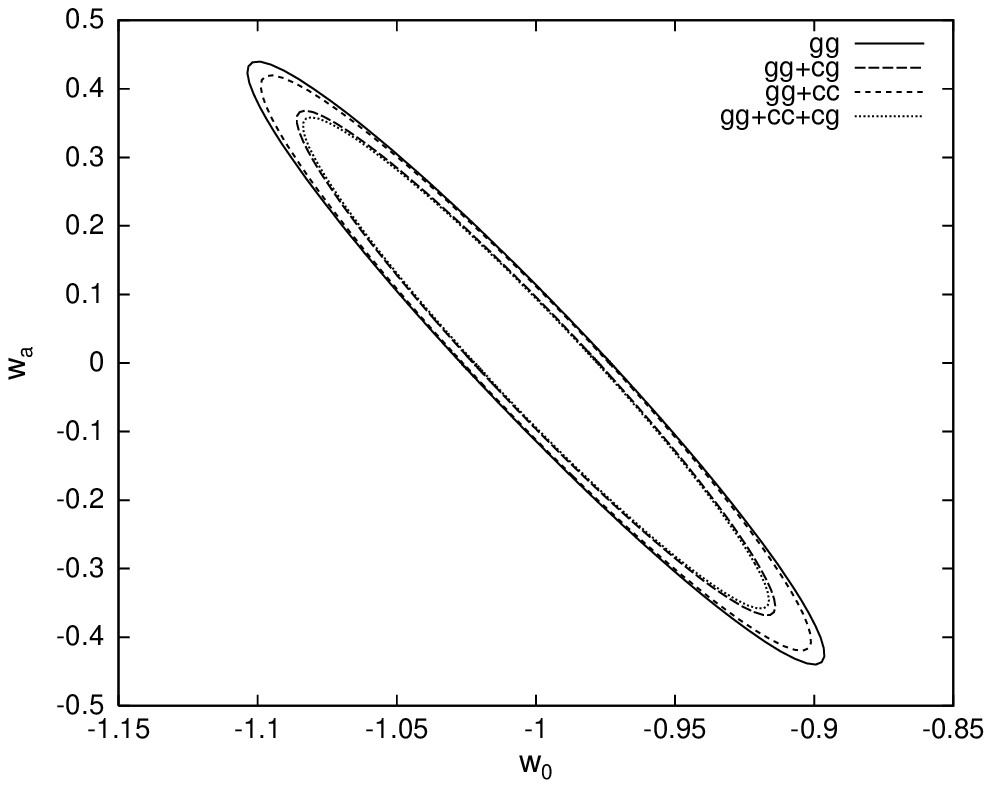}
\caption{1-$\sigma$ error contours on the $w_0$-$w_a$ plane. Solid, dashed, short-dashed, and dotted lines correspond to the $g-g$, $(g-g)+(c-g)$, $(g-g)+(c-c)$, and $(g-g)+(c-c)+(c-g)$ cases, respectively.}
\label{fig07}
\end{figure}

In Fig. \ref{fig06} we show in greater detail the improvement on the $m_g^{\low}$-$\alpha$ plane brought by the inclusion of the information from the cross-pairs. Although the constraints in the $c-g$ case are relatively weak a different degeneracy direction as compared to the $g-g$ case still helps in shrinking a $g-g$ error ellipse by a factor of $\sim 2$ (see Table \ref{tab1}). Also, as one of the main goals of these future redshift surveys is to put constraints on the properties of dark energy, Fig. \ref{fig07} presents the performance on the $w_0$-$w_a$ plane in greater detail. The dark energy FOMs we obtain are $26.9$, $38.5$, $28.8$, and $40.4$ for the $g-g$, $(g-g)+(c-g)$, $(g-g)+(c-c)$, and $(g-g)+(c-c)+(c-g)$ cases, respectively.\footnote{For the flat models with a scalar spectral index $n_s=1$ (as was assumed so far) by adding extra information obtainable from the CMB measurements by the {\sc Planck} satellite these FOMs increase to $81.9$, $125$, $86.1$, and $135$, respectively.} Here we see that the extra information from the cross-pairs leads to a relatively modest gain: error ellipse is compressed by a factor of $\sim 1.4$ as compared to the $g-g$ case taken separately. We also see that the additional gain provided by the inclusion of the cluster-cluster pairs is significantly smaller than that. This is also the case for the rest of the parameters, which is why we have not included $c-c$ case in Fig. \ref{fig05} and Table \ref{tab1}.    

\begin{figure}
\centering
\includegraphics[width=\plotwd]{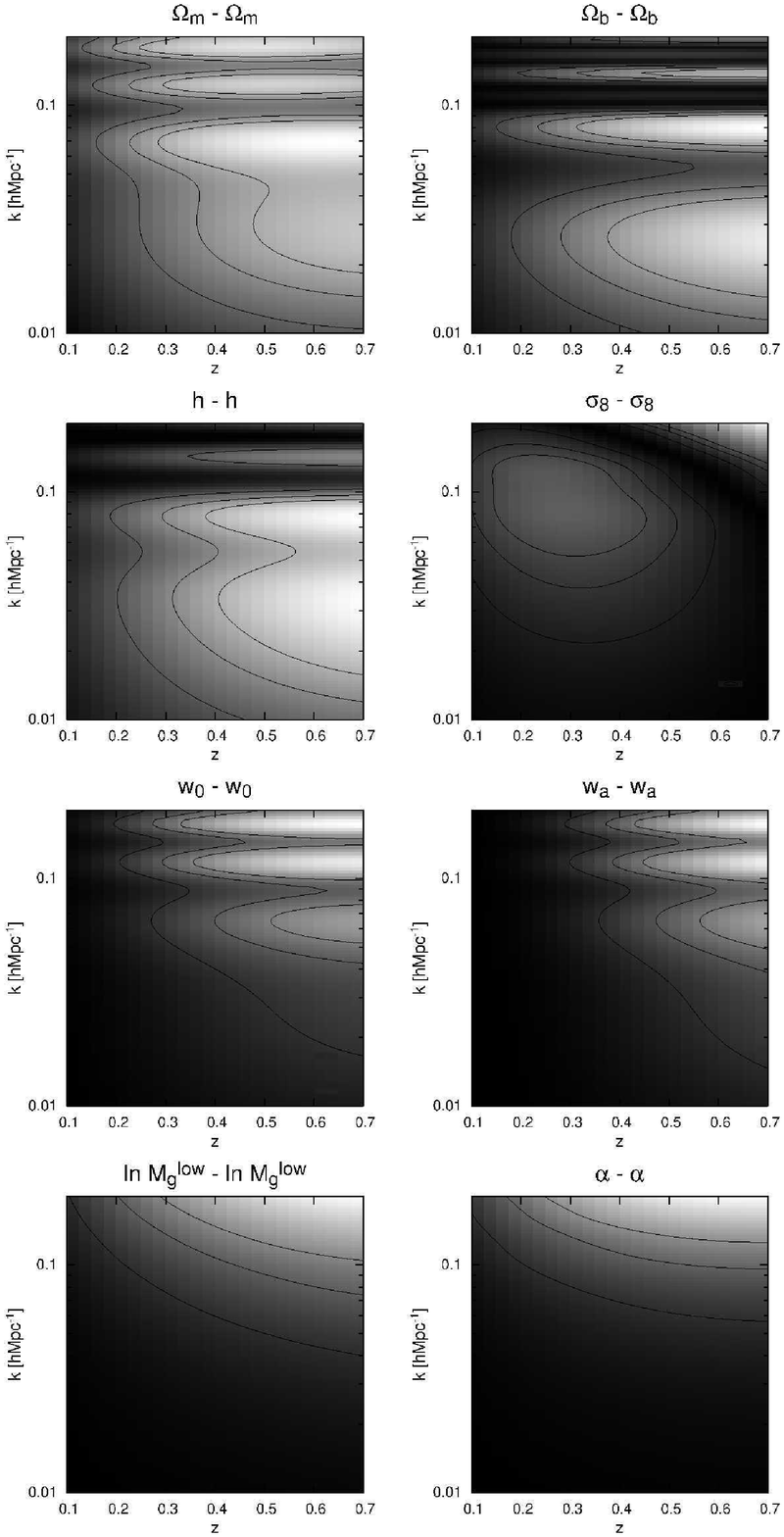}
\caption{Distribution of the $c-g$ Fisher information (for the diagonal matrix entries only) on the wavenumber-redshift plane, i.e. the total Fisher information is given by the integral over this plane.The discrete contour lines correspond to the isocontours enclosing $50$, $70$, and $90\%$ of the total Fisher information.}
\label{fig09c}
\end{figure}

In our Fisher matrix calculations we have focused only on a wavenumber interval $k=0.01-0.2\, h\,\rm{Mpc}^{-1}$. Due to formation of the nonlinear structures the Gaussianity assumptions is not adequate for wavenumbers larger than $\sim 0.2\, h\,\rm{Mpc}^{-1}$. At those small scales the information is transferred out of the two-point function to the higher order moments, increasing the variance of the power spectrum above it's Gaussian expectation, i.e. to calculate the variance one has to include also a contribution from the trispectrum (see e.g. \citealt{1999ApJ...527....1S}). Moreover, there have been a few studies (e.g. \citealt{2006MNRAS.371.1205R,2007MNRAS.375L..51N}) that indicate that there is not much information left in the power spectrum at those scales. As we have chosen to divide our survey volume into relatively narrow redshift shells it is clear that we are not able to probe the very large Fourier modes. However, the modes with wavenumbers above $0.01\,h\,\rm{Mpc}^{-1}$ are still relatively well sampled. We have checked that our results are rather insensitive to lowering the wavenumber boundary below it's assumed value of $k=0.01\,h\,\rm{Mpc}^{-1}$. For the $c-g$ case this can be seen by looking at Fig. \ref{fig09c} where we have shown how Fisher information (for the diagonal matrix entries only) is distributed on the wavenumber-redshift plane, i.e. the total Fisher information is given by the integral over this plane. There the discrete contour lines correspond to the isocontours enclosing $50$, $70$, and $90\%$ of the total Fisher information. One can see that for most of the parameters the surface drops off rather rapidly towards large scales and thus $90\%$ contour does not reach $k=0.01\,h\,\rm{Mpc}^{-1}$ line. For the $g-g$ and $c-c$ case the picture is qualitatively similar.    

\begin{table}
\caption{Figures-of-merit (FOMs), defined as the inverse areas of the 1-$\sigma$ error ellipses, for most of the free parameter combinations. The diagonal entries (e.g. $\Omega_m$-$\Omega_m$) give the inverse of the corresponding parameter's 1-$\sigma$ error interval, instead. In each cell the numbers from top to bottom correspond to $c-g$, $g-g$, $(g-g)+(c-g)$, and the ratio of $(g-g)+(c-g)$ and $g-g$ FOMs.}
\label{tab1}
\begin{tabular}{l||l|l|l|l|l|l|l|l}
 & $\Omega_m$ & $\Omega_b$ & $h$ & $\sigma_8$ & $w_0$ & $w_a$ & $\ln m_g^{\low}$ & $\alpha$ \\
\hline
\hline
            &\tiny{165} &\tiny{5610}&\tiny{673}&\tiny{506}  &\tiny{244} &\tiny{60.4} &\tiny{52.5}&\tiny{612}\\
$\Omega_m$  &\tiny{232} &\tiny{12,300}&\tiny{1510}&\tiny{1420}&\tiny{523}&\tiny{129} &\tiny{546} &\tiny{1670}  \\
            &\tiny{290} &\tiny{19,100}&\tiny{2570}&\tiny{3450}&\tiny{797}&\tiny{197} &\tiny{998} &\tiny{3470} \\
            &\bf{1.25}&\bf{1.55}  &\bf{1.70}&\bf{2.43}  &\bf{1.53}&\bf{1.53}&\bf{1.83}&\bf{2.07}  \\
\hline
            &     &\tiny{245} &\tiny{1550}  &\tiny{736}  &\tiny{328} &\tiny{73.2} &\tiny{78.3}&\tiny{970}  \\
$\Omega_b$  &-----&\tiny{384} &\tiny{4020}  &\tiny{2470}&\tiny{827}&\tiny{189} &\tiny{903} &\tiny{2950}  \\
            &     &\tiny{477} &\tiny{6580}&\tiny{5700}&\tiny{1250}&\tiny{282} &\tiny{1600} &\tiny{5850}\\
            &     &\bf{1.24}&\bf{1.64}  &\bf{2.31}  &\bf{1.51}&\bf{1.50}&\bf{1.77}&\bf{1.98}  \\
\hline
            &     &     &\tiny{26.9}&\tiny{87.2} &\tiny{34.0} &\tiny{7.81}&\tiny{9.49}&\tiny{127}  \\
$h$         &-----&-----&\tiny{43.7}&\tiny{328}&\tiny{88.7} &\tiny{20.9}&\tiny{107}&\tiny{372}  \\
            &     &     &\tiny{58.7}&\tiny{758}&\tiny{144} &\tiny{33.5}&\tiny{199}&\tiny{754} \\
            &     &     &\bf{1.34}&\bf{2.31}&\bf{1.62}&\bf{1.60}&\bf{1.86}&\bf{2.03} \\
\hline
            &     &     &     &\tiny{21.6}&\tiny{30.8} &\tiny{6.88}&\tiny{16.9}&\tiny{87.5}  \\
$\sigma_8$  &-----&-----&-----&\tiny{44.2} &\tiny{92.8} &\tiny{22.2} &\tiny{163} &\tiny{581} \\
            &     &     &     &\tiny{85.6} &\tiny{226} &\tiny{53.2} &\tiny{344} &\tiny{1260} \\
            &     &     &     &\bf{1.94}&\bf{2.44}&\bf{2.39}&\bf{2.12}&\bf{2.17} \\
\hline
            &     &     &     &     &\tiny{9.13}&\tiny{9.86}&\tiny{3.00}&\tiny{33.6}  \\
$w_0$       &-----&-----&-----&-----&\tiny{14.6}&\tiny{26.9}&\tiny{34.1}&\tiny{105}  \\
            &     &     &     &     &\tiny{17.6}&\tiny{38.5}&\tiny{59.9}&\tiny{211}  \\
            &     &     &     &     &\bf{1.21}&\bf{1.43}&\bf{1.76}&\bf{2.00} \\
\hline
            &     &     &     &     &     &\tiny{2.09}&\tiny{0.675}&\tiny{7.70} \\
$w_a$       &-----&-----&-----&-----&-----&\tiny{3.45}&\tiny{8.06}&\tiny{25.0} \\
            &     &     &     &     &     &\tiny{4.12}&\tiny{13.9}&\tiny{49.3}  \\
            &     &     &     &     &     &\bf{1.19}&\bf{1.73}&\bf{1.97} \\
\hline
                &     &     &     &     &     &     &\tiny{2.27}&\tiny{11.0} \\
$\ln m_g^{\low}$&-----&-----&-----&-----&-----&-----&\tiny{16.6}&\tiny{320}  \\
                &     &     &     &     &     &     &\tiny{24.2}&\tiny{652}  \\
                &     &     &     &     &     &     &\bf{1.46}&\bf{2.04} \\
\hline
            &     &     &     &     &     &     &     &\tiny{26.6} \\
$\alpha$    &-----&-----&-----&-----&-----&-----&-----&\tiny{52.1} \\
            &     &     &     &     &     &     &     &\tiny{86.4}  \\
            &     &     &     &     &     &     &     &\bf{1.66}
\end{tabular}
\end{table}

\section{Conclusions and Discussion}
There are several wide field galaxy and cluster surveys planned for the nearest future. In the simplest approach one would analyze these independently, thus neglecting the extra information provided by the cluster-galaxy cross-pairs. In this paper we have focused on the possible synergy between these surveys by investigating the amount of information encoded in the cross-pairs. Since in our study we focus on two-point clustering statistics our approach is complementary to the number count analysis, which in its usual form assumes validity of the Poissonian statistics. \footnote{In case of the galaxy cluster samples the complementarity of the number count and clustering analysis has been investigated by \citet{2004ApJ...613...41M}. For example, these authors illustrate that combining number counts and $c-c$ power spectrum helps in reducing the error bar on $\sigma_8$ by a factor of two for an SPT-like survey. This is similar to the improvement on $\sigma_8$ we obtain by combining $c-g$ with $g-g$. Due to the sparseness of the cluster sample the additional inclusion of the $c-c$ power leads to only mildly stronger constraints. To evaluate the absolute error on $\sigma_8$ one should combine our pair statistics with the cluster number counts, which is beyond the scope of the current work.}

To model the cluster-galaxy cross-spectrum we have used the Halo Model framework. We have carried out a Fisher matrix analysis for a BOSS-like galaxy redshift survey targeting luminous red galaxies over $10,000$ deg$^2$ of sky and over the redshift range of $z=0.1-0.7$, and a hypothetical mass-limited cluster redshift survey (with a threshold mass $m_c^{\low}=1.7\times10^{14}\,h^{-1}M_{\odot}$) over the same volume. For simplicity we have assumed spatially flat models with the usual $w_0-w_a$ parametrization for the dark energy equation of state. For the HOD we have used a simple description given by Eqs. (\ref{eq9}),(\ref{eq10}),(\ref{eq11}). Since clustering does not depend on $m_0$ we are left with two HOD parameters: $\alpha$ and $m_g^{\low}$. Adding to these $m_c^{\low}$ and six cosmological parameters: $\Omega_m$,$\Omega_b$,$h$,$\sigma_8$,$w_0$, and $w_a$, we have in total nine free parameters. 

As on small scales the cluster-galaxy cross-spectrum probes directly density profiles of the halos, instead of the profile convolved with itself, as is the case for the galaxy power spectrum, we are sensitive to a different combination of the HOD parameters. This leads to stronger constraints on the HOD parameters $\alpha$ and $m_g^{\low}$ as compared to the constraints obtainable from the galaxy data taken separately, e.g. the inclusion of the cross-pairs leads to a factor of $\sim 2$ compression of the error ellipses on the $m_g^{\low}$-$\alpha$ plane. Also, the extra information in the cross-pairs leads to a factor of $\sim 2$ tighter constraint on $\sigma_8$. The best performance is obtained on $\sigma_8$-$\Omega_m$, $\sigma_8$-$w_0$, and $\sigma_8$-$w_a$ planes where the error ellipses get reduced by a factor of $\sim 2.4$. 

For the dark energy equation of state parameters $w_0$ and $w_a$ the improvement is somewhat weaker: inclusion of the cluster-galaxy cross-pairs leads to a factor of $\sim 1.4$ increase in the dark energy figure-of-merit as compared to the galaxy-galaxy case taken separately. For the rest of the results see Fig. \ref{fig05} and Table \ref{tab1}.

It is worth mentioning that the small scale information in the one-halo term of the cluster-galaxy power spectrum is essentially equivalent to the information gained by directly fitting for the galaxy distribution within the halos. $P^{1h}_{c-g}$ is just a Fourier transform of the weighted mean of the halo density profiles. From the rather sparse data provided by the assumed redshift surveys one cannot reliably perform profile fitting on object by object basis, and thus probably has to rely on stacking methods, which also lead to some sort of a weighted profile. Thus these two approaches should give essentially the same results.

In our analysis we have assumed that the threshold cluster mass $m_c^{\low}$ can be calibrated with an accuracy of $10\%$. We find that even without the assumed $10\%$ prior for $m_c^{\low}$ the information encoded in the $c-g$ pairs is able to self-calibrate $m_c^{\low}$ to an accuracy of $\sim 4\%$, and thus our results are rather insensitive to this additional prior. On the other hand, treating $m_c^{\low}$ as a known quantity and fixing it to the fiducial value of $1.7\times10^{14}\,h^{-1}M_{\odot}$ we obtain $\sim 5$ times tighter constraint on $\sigma_8$. Similarly, all the error ellipses involving $\sigma_8$ as one of the parameters get reduced by factors of $6-10$ in this case, the most remarkable being a factor of $\sim 10$ reduction of the error ellipse on the $\sigma_8$-$m_g^{\low}$ plane. For the other parameters the precise knowlwdge of $m_c^{\low}$ leads to much milder improvements, e.g. $m_g^{\low}$-$\alpha$ and $w_0$-$w_a$ error ellipses now decrease by factors of $\sim 3$ and $\sim 1.5$, respectively.

In our analysis we have simply added Fisher matrices corresponding to the galaxy-galaxy and cluster-galaxy pairs. This is a valid approach once the density field is not over-sampled. If that is not the case, the extra pairs do not help in increasing the information, i.e. then we cannot simply add Fisher matrices. However, in our case the Fisher information comes mostly from the larger wavenumbers as there are simply more high-$k$ modes available (Fig. \ref{fig09c} shows this for the $c-g$ case). As can be seen from Fig. \ref{fig03}, for large wavenumbers our hypothetical surveys are surely not over-sampled, and thus simply adding the Fisher matrices seems to be a valid approximation.    

As a final comment we point out that the formalism as presented in Sec. \ref{sec2} can also be applied to the photometric redshift surveys such as DES, Pan-STARRS, DUNE, and LSST. In this case including cluster-galaxy pairs might be very beneficial since the photo-z errors for the clusters are usually significantly smaller than that for the typical galaxies.
\bibliographystyle{aa}
\bibliography{references}

\begin{appendix}
\section{Power spectra in photometric redshift-space}\label{appb}
As there will be several large imaging surveys performed in the future (e.g. DES, Pan-STARRS, DUNE, LSST), we present here a minimal extension of the formalism for the case when only photometric redshifts are available.  
For simplicity we assume that the conditional probability distribution for the spectroscopic redshift given the photometric one $P(z_{\rm spec}|z_{\rm photo})$ is a Gaussian with a mean value of $z_{\rm photo}$ (i.e. we assume zero bias) and with dispersion $\delta z$. This leads to the extra damping factors for the density contrast, which have the same form as in Eq. (\ref{eq19}) with the 1D spatial smoothing scale $\sigma$ given as $\sigma=\frac{c}{H_0}\delta z$. We denote the corresponding smoothing scales for clusters and galaxies as $\sigma_c^{\rm photo}$ and $\sigma_g^{\rm photo}$, respectively. As for any realistic case photo-z errors change with redshift, $\sigma_c^{\rm photo}$ and $\sigma_g^{\rm photo}$ are modeled as functions of $z$.

Thus for the dark matter we have an effective smoothing scale $\sigma_m(m,z)$, for the galaxies $\sigma_g(m,z)\equiv\sqrt{(\sigma_m(m,z))^2 + (\sigma_g^{\rm photo}(z))^2}$ (FOG + photo-z errors), and for the clusters $\sigma_c(z)\equiv\sigma_c^{\rm photo}(z)$ (photo-z errors only). We also define the quantity $\sigma_{c-g}(m,z)\equiv\sqrt{(\sigma_c(z))^2 + (\sigma_g(m,z))^2}=\sqrt{(\sigma_m(m,z))^2 + (\sigma_c^{\rm photo}(z))^2 + (\sigma_g^{\rm photo}(z))^2}$. Having introduced these new smoothing scales we can obtain results valid for the photometric redshift-space if we substitute $\sigma_m(m,z)$ with $\sigma_g(m,z)$ in Eqs. (\ref{eq24}) and (\ref{eq33}), $\sigma_m(m,z)$ with $\sigma_{c-g}(m,z)$ in Eq. (\ref{eq25}), and include an additional factor of $\mathcal{F}\left[k\sigma_c(z)\mu\right]$ in Eq. (\ref{eq32}).

\section{Covariances and Fisher matrices}\label{appa}
Under the assumption of Gaussianity the power spectrum covariance matrix can be expressed as \citep{1994ApJ...426...23F, 1997PhRvL..79.3806T}:
\begin{equation}  
^x\mathbf{C}_{mn}\simeq \frac{2 P_x^2(k_n)}{V_n V_{{\rm eff}}^x(k_n)}\delta_{mn},\,\label{eq38}
\end{equation}
where $P_x(k_n)$ denotes the power spectrum of clusters/galaxies for the wavenumber bin centered at wavenumber $k_n$, $V_n \equiv 4\pi k_n^2 \Delta k_n/(2\pi)^3$ ($\Delta k_n$ -- width of the wavenumber bin) is the volume of the shell in Fourier space, and the effective volume $V_{{\rm eff}}^x(k)$ is given as:
\begin{equation}  
V_{{\rm eff}}^x(k) \equiv \int \left[ \frac{\bar{n}_x(\mathbf{r})P_x(k)}{1 + \bar{n}_x(\mathbf{r})P_x(k)}\right]^2\dd \mathbf{r} = \left[ \frac{\bar{n}_x P_x(k)}{1 + \bar{n}_x P_x(k)}\right]^2 V\,,\label{eq39} 
\end{equation}
where $V$ is the total volume of the survey. Here the second equality for $V_{{\rm eff}}(k)$ is valid only if the selection function of the cluster/galaxy sample, $\bar{n}_x(\mathbf{r})$, is independent of position. 

Following the lines of thought in Appendix B of \citet{1994ApJ...426...23F}, which lead to the result given in Eq. (\ref{eq38}) we find that the similar result for the cluster-galaxy cross-spectrum can be expressed as:
\begin{equation}  
^{c-g}\mathbf{C}_{mn}\simeq \left[ \frac{P_{c-g}^2(k_n)}{V_n V} + \frac{P_{c-c}(k_n)P_{g-g}(k_n)}{V_n V_{{\rm eff}}^{c-g}(k_n)} \right]\delta_{mn},\,\label{eq40}
\end{equation}
where the effective volume $V_{{\rm eff}}^{c-g}(k)$:
\begin{eqnarray}
V_{{\rm eff}}^{c-g}(k) &\equiv& \int \left[ \frac{\bar{n}_c(\mathbf{r})P_{c-c}(k)}{1 + \bar{n}_c(\mathbf{r})P_{c-c}(k)}\right]\left[ \frac{\bar{n}_g(\mathbf{r})P_{g-g}(k)}{1 + \bar{n}_g(\mathbf{r})P_{g-g}(k)}\right] \dd \mathbf{r} \nonumber \\ &=& \left[ \frac{\bar{n}_c P_{c-c}(k)}{1 + \bar{n}_c P_{c-c}(k)}\right] \left[ \frac{\bar{n}_g P_{g-g}(k)}{1 + \bar{n}_g P_{g-g}(k)}\right] V\,.\label{eq41} 
\end{eqnarray}
The first term in Eq. (\ref{eq40}) is the analog of Eq. (\ref{eq38}) with the difference of the missing factor of two and also that here instead of the effective volume we have a survey volume $V$, which is due to the fact that in case of the cross-spectrum we do not have a shot-noise term raising from self-pairs.
To find the corresponding Fisher matrices we follow the derivation in \citet{1997PhRvL..79.3806T}. In our calculations we divide the survey volume into relatively narrow redshift bins. This allows us to approximate the spatial number densities of objects as constant quantities within the bins, i.e. we can use the second equalities of Eqs. (\ref{eq39}) and (\ref{eq41}). 

The Fisher matrices for the anisotropic cluster/galaxy spectra can be given as:  
\begin{eqnarray}
^{x}\mathbf{F}_{ij}(z_n)&\simeq& \frac{V(z_n)}{8\pi^2} \cdot \int\limits_{-1}^{1} \dd \mu \int\limits_{k_{\min}}^{k_{\max}}\dd k \,k^2 w^x(k,\mu|z_n)\nonumber\\ &\cdot& \frac{\partial \ln P_x(k,\mu|z_n)}{\partial \Theta_i}\cdot\frac{\partial \ln P_x(k,\mu|z_n)}{\partial \Theta_j}\,,\label{eq42}\\
w^x(k,\mu|z_n) &\equiv& \left[ \frac{\bar{n}_x(z_n) P_x(k,\mu|z_n)}{1 + \bar{n}_x(z_n) P_x(k,\mu|z_n)} \right]^2 = \frac{V_{{\rm eff}}^x(k,\mu|z_n)}{V(z_n)}\,.\label{eq43}
\end{eqnarray}
Here $z_n$ denotes the central redshift and $V(z_n)$ the comoving volume corresponding to the $n$-th redshift bin. 

The corresponding results for the cluster-galaxy cross-spectrum are given as:
\begin{eqnarray}
^{c-g}\mathbf{F}_{ij}(z_n)&\simeq& \frac{V(z_n)}{8\pi^2} \cdot \int\limits_{-1}^{1} \dd \mu \int\limits_{k_{\min}}^{k_{\max}}\dd k \,k^2 w^{c-g}(k,\mu|z_n)\nonumber\\ &\cdot& \frac{\partial \ln P_{c-g}(k,\mu|z_n)}{\partial \Theta_i}\cdot\frac{\partial \ln P_{c-g}(k,\mu|z_n)}{\partial \Theta_j}\,,\label{eq44}\\
w^{c-g}(k,\mu|z_n) &\equiv& \frac{2}{ 1 + \frac{\left[ 1 + \bar{n}_c(z_n) P_{c-c}(k,\mu|z_n)\right]\left[ 1 + \bar{n}_g(z_n) P_{g-g}(k,\mu|z_n) \right]}{\bar{n}_c(z_n) \bar{n}_g(z_n) P_{c-g}^2(k,\mu|z_n)}}\,.\label{eq45}
\end{eqnarray}
The total Fisher matrices are then the sums over the ones corresponding to different redshift bins:
\begin{equation}
^x\mathbf{F}_{ij} = \sum \limits_n {^x\mathbf{F}_{ij}}(z_n)\,.
\end{equation}
In the above relations $\Theta_i$ represents the $i$-th component of the parameter vector $\mathbf{\Theta}$, which in our case consists of cosmological plus HM parameters. According to the Cram\'er-Rao inequality there does not exist an unbiased method that can measure the $i$-th parameter $\Theta_i$ with error bars less than $1/\sqrt{\mathbf{F}_{ii}}$. If the other parameters are not known but estimated from the data as well, the minimum achievable error is given by $\sqrt{\mathbf{F}^{-1}_{ii}}$ instead. 

\end{appendix}

\end{document}